\documentclass[12pt,a4paper]{article}

\usepackage{amsmath}

\usepackage{amsfonts}
\usepackage{amssymb}

\begin{document}

\title{Quantum Information Biology: from information interpretation of quantum mechanics to applications in molecular biology and 
cognitive psychology}

\author{Masanari Asano\\
Liberal Arts Division, Tokuyama College of Technology, Japan\\
 Irina Basieva, Andrei Khrennikov\\
 International Center for Mathematical Modeling \\in Physics and Cognitive Science \\
Linnaeus University,  V\"axj\"o, Sweden \\
Masanori Ohya and Yoshiharu Tanaka \\
Department of Information Sciences\\Tokyo University of Science, Japan\\
Ichiro Yamato\\Department of Biological Science and Technology\\ Tokyo University of Science, Japan}

\maketitle

\begin{abstract}
We discuss foundational issues of  quantum information biology (QIB) -- one of the most successful  applications of the quantum formalism outside of physics. 
QIB provides a multi-scale model of information processing in bio-systems: from proteins and cells to cognitive and social systems. 
This theory has to be sharply distinguished from ``traditional quantum biophysics''. The latter is about quantum bio-physical processes, e.g., in cells or brains.
QIB models the dynamics of information states of bio-systems. It is based on the quantum-like paradigm: complex bio-systems process information in accordance 
with the laws of quantum information and probability. This paradigm is supported by plenty of statistical bio-data collected at all scales, from molecular biology and 
genetics/epigenetics to cognitive psychology and behavioral economics. We argue that the information interpretation of quantum mechanics 
(its various forms were elaborated by Zeilinger and Brukner, Fuchs and Mermin, and D' Ariano)  is the most natural interpretation of QIB. We also point out that QBIsm (Quantum Bayesianism) can serve 
to find a proper interpretation of bio-quantum probabilities. Biologically QIB is based on two principles: a) adaptivity; b) openness (bio-systems are fundamentally open).
These principles are mathematically represented in the framework of  a novel formalism -- quantum adaptive dynamics which, in particular, contains the standard theory of 
open quantum systems as a special case of adaptivity (to environment). 
\end{abstract}

keywords: quantum biological information, quantum adaptive dynamics, open quantum systems, information interpretation, QBism, molecular biology, genetics, cognition

\section{Introduction}

In this paper we discuss a novel field of research, {\it quantum information biology} (QIB). It is based on application of 
a novel quantum information formalism \cite{a1}--\cite{BOOK},  {\it  quantum adaptive dynamics}, outside 
of physics.\footnote{QIB is not about quantum physics of bio-systems (see \cite{Arndt} for extended review), in particular, not about 
quantum physical modeling of cognition, see section \ref{PP} for details. Another terminological possibility would be ``quantum bio-information''. However, the latter term, bio-information,
has been reserved for a special part of biological information theory studying computer modeling of sequencing of DNA. We do not want to be 
associated with this activity having totally different aims and mathematical tools.} 
Quantum adaptive dynamics describes in the most general setting mutual adaptivity of information states of systems of any origin 
(physical, biological, social, political) as well as mutual adaptivity their co-observations, including self-observations (such as performed by the brain).

Nowadays {\it quantum information theory} is widely applied to quantum computers, simulators, cryptography, and teleportation. 
Quantum information theory is fundamentally based  on quantum probability (QP). 

We noticed that biological phenomena often violate the basic laws of classical probability (CP)  \cite{34}--\cite{30}; in particular, the law of total probability (LTP). 
Therefore in biology (treated widely  as covering cognition and its derivatives: psychology and decision making, sociology and behavioral economics and finances) 
it is natural to explore one of the most advanced and well known nonclassical probability theories, namely, QP. 
This leads to the quantum information representation of  biological 
information flows which differs crucially from the classical information representation. 

Since {\it any bio-system $S$ is fundamentally open} (from a cell to a brain or a social system), 
i.e., it cannot survive without contacts with environment (physical, biological, social), 
it is natural to apply the powerful and well developed apparatus of theory of open quantum systems\cite{OV} 
to the description of biological information flows. Adaptivity of the information state of $S$ to an environment is a special form 
of bio-adaptivity: mutual adaptivity of information states of any pair of bio-systems $S_1$ and $S_2.$ Since quantum information theory 
provides a very   general description of information processing, it is natural to apply it to model adaptive dynamics. This led us to 
quantum adaptive dynamics. Surprisingly we found that such an {\it operational quantum formalism} can be applied to any biological scale 
by representing statistical experimental data by means of quantum probability and information.
During the last years a general model representing all basic information flows in biology (from molecular biology to cognitive science and psychology and to evolution) 
on the basis of quantum information/adaptive dynamics was elaborated. In a series of our works \cite{a1}--\cite{BOOK}
 the general scheme of embedding of biological information 
processing into the quantum information formalism was presented and the foundational issues related to usage of quantum representation for macroscopic bio-systems, 
such as genome, protein,..., brain,..., bio-population were discussed in details and clarified.

Our theory can be considered as the informational basis of {\it decision making.} Each bio-system (from a cell to a brain and to a social or ecological system) 
is permanently in process of decision making. Each decision making can be treated as a self-measurement or more generally as an adaptive reply to signals 
from external and internal environments.

On this basis, we believe that QIB is the most predictive tool 
know our future state on earth. We expect that {\it this quantum-like operator formalism is a kind of brave trial to unify our social and natural sciences.}  

\section{Interrelation of quantum bio-physics and information biology}
\label{PP}

This section is devoted to comparison of quantum bio-physics with QIB. Those who have been satisfied 
with a brief explanation given in footnote 1 can jump over this section.  

First of all, we emphasize that our applications of the methods of QM to modeling of information flows in bio-systems  
have no direct relation to {\it physical quantum processes} in them, cf. \cite{Arndt}. For example, by considering quantum information processing
by a cell or by a protein molecule \cite{a1}--\cite{BOOK} we do not treat them as quantum {\it physical systems.} In fact, already a single cell or even 
a protein molecular are too big to be treated in the conventional QM-framework which was elaborated for microscopic systems. (N. Bohr emphasized 
the role of the fundamental quantum of action - the Planck  constant $h.$) Of course, in the quantum foundational community so-called 
macroscopic quantum phenomena have been discussed a lot. However, it seems that magnitudes of some important physical parameters of bio-systems do not match 
with scales of macro-quantum phenomena (e.g., Tegmark's \cite{Tegmark} critique of ``quantum mind'', cf. with \cite{Arndt}). For example, bio-systems are too hot comparing with Bose-Einstein condensate or Josephson-junction. At the same time 
we can point to experiments of A. Zeilinger continued by  Arndt, see  \cite{Arndt}, on interference for macro-molecules, including viruses. They were performed at high temperature. 
We can also point to recent studies claiming the quantum physical nature of the photo-synthesis. 

In any event in QIB  we are not interested in all complicated problems of macro- and, in particular, bio-quantumness.
In particular, by applying our formalism to cognition we escape involvement in hot debates about a possibility to treat the hot and macroscopic brain 
as processing quantum physical  (entangled) states, since we do not couple our model to the ``quantum physical brain project'', e.g., 
theories of  Penrose \cite{P1} and Hameroff \cite{H1}. For example, the complex problem whether micro-tubules in the brain are able 
to preserve quantum physical entanglement (before its decoherence) for a sufficiently long time, e.g., to perform a step of quantum computation,
is outside of our theory.  QIB describes well processing of mental informational states, independently of who will be right:
Penrose and Hameroff or Tegmark. 
 (It is also possible that this debate will continue
forever.)  The ability of a bio-system to operate with superpositions of information states 
including brain's ability to form superpositions of mental states is the key-issue. 

For a moment, there is no commonly accepted model of creation of superposition of information states by a bio-system, in particular, 
there is no a proper neurophysiological model of creation of mental superposition.
For the latter, we can mention a series of attempts to model states superposition with the aid of classical oscillatory processes in the brain, 
as it was proposed by J. Acacio de Barros and P. Suppes \cite{1}, 
including classical electromagnetic waves in the brain, see, e.g., for a model  \cite{KHRBRAIN} proposed by A. Khrennikov. 
In these studies mental states are associated with classical 
physical waves or oscillators, the waves  discretized with the aid of thresholds induce the probabilistic interference 
exhibited in the form of violations of LTP.

In general  we consider violations of  LTP 
by  statistical data collected in bio-science, from molecular biology to cognitive psychology, as confirmation of the ability of  
bio-system  to operate  with states superpositions. In an experiment, in QM as well as in biology, we do not detect superposition of  ``individual information states''
such as superposition of two classical waves. {\it Informational states of bio-systems are probabilistic amplitudes.} In QIB 
we discuss the ability of bio-systems to represent probabilities by vector amplitudes and operate with such amplitudes by using the operations which are mathematically 
represented as the matrix calculus. From this operational viewpoint an electron does not differ so much from a cell, or a brain, or a social system.  

In  the series of works, see \cite{UB} for references,  it was shown  that, for probabilistic data of any origin, violation of LTP leads to representation of states by 
probability amplitudes  (usual complex  or more general, so-called hyperbolic)- {\it the constructive wave function approach.} It might be that bio-systems really 
use this algorithm of production  of amplitudes from probabilities -- {\it the quantum-like representation algorithm.} In a bio-system probability can be 
treated as frequency. And frequencies can  (but need not)  be encoded in oscillatory processes in bio-systems.  

Finally, we speculate that the brain could, in principle, use a part of its {\it spatial representation} ``hardware'' (its deformation in the process of evolution)
for the vector representation of probabilities and hence, for decision making and processing of information at the advanced level. For a moment, this is a pure speculation. 

\section{From information physics to information biology}

\subsection{Operational approach} 

In QIB, similarly to quantum physics\footnote{N. Bohr always pointed out that quantum theory describes the results of measurements and emphasized
the role of an observer; he stressed that the whole experimental arrangement has to be taken into account; W. Heisenberg and W. Pauli had similar views, 
see A. Plotnitsky \cite{PL1}--\cite{PLX} for detailed analysis of their views.},
we  treat the quantum formalism as an {\it operational formalism} describing 
(self-)measurements performed by bio-systems.   

Neither QM nor biology (in particular, cognitive science) can explain why systems under study produce such random outputs - violating the CP-laws.   
Moreover, in QM, according  to the Copenhagen interpretation, it is in principle impossible to provide
some ``explanation'' of quantum random behavior, e.g., by using a more detailed description with the aid of so-called {\it hidden variables.} In spite of this explanatory 
gap, QM is one of the most successful scientific theories. One may hope that a similar operational approach would finally lead to creation of a novel and fruitful 
theory  for bio-systems.  

We remark that in recent studies of D' Ariano et al. \cite{dariano}--\cite{darianoX} the quantum formalism  was derived from a set of purely operational postulates reflecting features 
of a very general measurement scheme. Such an approach can be applied to QIB as a theory of (self-)measurements. However, one should carefully 
check matching of D' Ariano's postulates with decision making done by bio-systems (considered as self-measurements). It is natural to expect that some of these (physically natural) postulates 
have to be  modified and that new postulates can be in use. This is an interesting project waiting for its realization.

\subsection{Information interpretation of quantum mechanics}

We also emphasize that recently the {\it informational interpretation} of the quantum state started 
to play an important role in QM and, especially, in quantum information theory, see the works of A. Zeilinger and C. Brukner \cite{Z1}--\cite{ZX}, C. Fuchs and D. 
Mermin \cite{F1}--\cite{FX},
and M. D' Ariano et al \cite{dariano}--\cite{darianoX}. By this interpretation the QM formalism is about the information processing related to experiments. 
The wave function is not treated as a physical state of, e.g., an individual electron. It is treated as representing information  
about possible outcomes of experiments. This viewpoint matches well with Schr\"odinger's treatment of the wave function 
as a table of expectations about possible outcomes.\footnote{We remark that Schr\"odinger's views on possible interpretations 
of the wave function changed crucially a few times. At the very beginning he considered the wave function as a new physical field. Then 
he pointed out that for a system of two electrons their wave function can be represented as a physical field  defined not on the physical three dimensional 
space, but on the six dimensional configuration space. At that time multi-dimensional spaces were considered as exotics, cf. with the modern situation 
in string theory. Therefore he gave up with the physical interpretation of quantum states. 
By discovering unusual  features of entangled states he concluded that such states can only be understood if they are treated as information states.
We remark that later, in 50th, Schr\"odinger stimulated by the success of quantum field theory tried again to come back to the physical field interpretation 
of the wave function.} The adherents of the information interpretation also often refer to the views of N. Bohr and W. Heisenberg who treated QM 
operationally. However, here the situation is complicated, since, for example, Bohr's views on the interpretation of the wave function changed 
many times during his life, see, for example, A. Plotnitsky for discussions \cite{PL1}-\cite{PLX}.

\subsection{QBism: Quantum Bayesian interpretation of quantum probability}

In its extreme subjective form the information interpretation is presented as QBism, {\it quantum Bayesian approach} -- C. Fuchs and D. Mermin, see, for example, \cite{F1}--\cite{FX}.
 QBists really  belief that the QM-formalism is simply a mathematical tool to assign subjective probabilities to expected outcomes of  experiments. QBism's position 
is very supporting for the part of  QIB related to the project ``quantum(-like) cognition''. Probabilities assigned to possible 
events by decision makers are subjective. It is worse with our more general project.  Subjective probability,  what would it mean for a cell? for a protein?
for a social or ecological system? 

Even for quantum cognition, we cannot simply (as C. Fuchs does in QM) assign a subjective probability to some event: ``the brain produces it in some way''.
By discussing the cognitive component of decision making it would be natural to try at least to imagine how a subjective probability is assigned to the concrete event.

One can speculate that, in fact, the probability which appears at the conscious level as the feeling of  ``subjective probability'' is an esemble probability 
at the unconscious level. Unconscious has a plenty of factors for and against the possibility of occurrence of some event $A.$ These factors are weighted 
by unconscious, so in the mathematical terms it operates with a probability measure corresponding to summation of weights of the factors in favor of $A.$\footnote{     
This unconscious-ensemble viewpoint to subjective probabilities in QM and in this way connection ``quantum-mental'' as the basis of QBism 
was advertised by one of the coauthors (AKH). During many years  it was discussed with C. Fuchs, but  his reaction was not supporting. Moreover, another author 
of this paper (IB) strongly rejects the unconscious-ensemble viewpoint on quantum probabilities which she otherwise interprets as subjective. The reader can see 
how difficult is the probability interpretation problem and that even keeping rigidly to the subjective interpretation one cannot say that the interpretation problem 
was solved completely.}

We remark that at the very beginning A. Zeilinger and C. Brukner did not emphasize so much the subjective probability viewpoint to QP, they stressed
that QM is about information, that this is a step towards, so to say, information physics. However,  the recent talk of A. Zeilinger 
at the conference ``50 years of Bell theorem'', Vienna, June 2014, was very much in the QBism spirit.

\subsection{Difference between the information interpretations of Zeilinger-Brukner and Fuchs-Mermin: the role of irreducible quantum randomness}

Although, as  we can see from the aforementioned lecture of Zeilinger, views of  Zeilinger-Brukner and Fuchs-Mermin are sufficiently close, 
it is important to point to one fundamental difference in their approaches. The interpretation of Zeilinger-Brukner is heavily based on the assumption 
that {\it quantum randomness is irreducible.} This idea that quantum randomness crucially differs from classical one was present already in von Neumann's book \cite{VN}.
He pointed out that quantum randomness is individual, e.g., even an individual electron is intrinsically random, and classical randomness is of the ensemble nature.
In a series of papers \cite{Z1}--\cite{ZX} this ideology was developed in the quantum information framework and it serves as the base of  the Zeilinger-Brukner interpretation.
It seems that QBism can proceed even without referring to the irreducible quantum randomness. 

In principle, in QIB we can proceed as QBists, i.e., without looking for biological sources of irreducible randomness.  On the other hand, it is important to pose the 
following stimulating question:

\medskip

{\it Is an individual bio-system intrinsically random?}
 
\subsection{Free will problem}

In principle, for human beings (and applications of QIB to cognition and decision making), one can try to couple the intrinsic quantum(-like) randomness
with {\it free will.}  In the recent debates on non/deterministic nature of quantum processes, the role of non/existence  of free  was actively 
discussed. At the moment two opposite positions aggressively coexist. We remark that the rejection of free will, the position of G. `t Hooft \cite{TH} known as super-determinism,
opens a possibility to treat QM as a deterministic theory, e.g. \cite{TH1}, \cite{TH2}. (We also remark that a part of quantum foundation community thinks that the issue of non/existence of free 
will is irrelevant to the problem of quantum (in)determinism.) 

Exploring the free will assumption may be helpful to motivate the presence of {\it intrinsic mental randomness} and in this way to couple QIB even closer 
to Zeilinger-Brukner interpretation. However, by proceeding in this way one has to be ready to meet a new problem of the following type:

\medskip

{\it Have a cell, DNA or protein molecular a kind of free will?}
 
\medskip   
  
We can formulate the problem even another way around:

\medskip

{\it Is human free will simply a special exhibition of intrinsic randomness which is present at all scales nature?}
 
\medskip

We remark that in general the problem of interrelation of classical and quantum probabilities and randomnesses is extremely complex. In some way it is easier 
for probabilities, since here we have satisfactory mathematical models, e.g., the Kolmogorov model \cite{K} (1933) of classical probability. However, even in the classical 
case a commonly acceptable notion of randomness has not yet been elaborated, in spite tremendous efforts of the during the last one hundred years, see \cite{INT}, 
\cite{arXiv0} for details.  

\subsection{Bohmian mechanics on information spaces and mental phenomena}
 
In fact, this position in application to mental phenomena and their unification with natural phenomena was expressed in a series 
of works, see, e.g.,   \cite{pi}, of one of the coauthors of this paper (A. Khrennikov), where a version of Bohmian mechanics on information spaces was in use. 
We remark that  some Bohmians, e.g., D. Bohm by himself and B. Hiley, interpreted the wave function as a field of information. But this information 
field was still defined on the physical space, because, for D. Bohm (as well as his predecessor L. De Broglie), the physical space was undiscussable reality.
In works of A. Khrennikov mental information fields were defined on ``information space'' reflecting hierarchic tree-like represenattion of information 
by cognitive systems (mathematically such mental trees can be represented with the aid of so called p-adic numbers \cite{pia}). However, later A. Khrennikov became 
less addicted to the idea of information reality. 

{\it The main problem of the information interpretation of QM is to assign a proper meaning to the notion of 
information.} It is very difficult to do -- especially by denying reality; one has to define information not about some real objectively existing stuffs,
but information as it is by itself. Of course, one can simply keeps to Zeilinger's position (private discussions): information is a primary notion 
which cannot be expressed through other ``primitive notions''. However, not everybody would be satisfied by such a solution of this interpretational problem. 
        
\subsection{Information interpretation as a biology friendly interpretation of quantum state}         
        
In any event the information interpretation matches perfectly with our aims, although its creators may be not support the attempts to apply it outside of physics.
(We remark that the position of A. Zeilinger was quite positive\footnote{In a series of private discussions with A. Khrennikov he expressed satisfaction 
that by exploring ideas of QM for modeling of cognition one finally breaks up the realistic attitude dominating in cognitive science.
At the same time he did not find reasons to use {\it precisely the quantum formalism} to model cognition: novel operational formalisms 
representing information processing might be more appropriate, see also section \ref{why?}.}.) 
 There is a chance  that physicists using this interpretation would welcome the born of quantum cognition and more generally QIB.
There was no any chance to find understanding of physicists keeping to the orthodox  Copenhagen interpretation -- the wave function as the most complete 
representation of the {\it physical state} of an individual quantum system such as an electron.

\section{Nonclassical probability? Yes! But,  why quantum probability?}
\label{why?}

One may say: ``Yes, I understood that the operational description of information processing in complex bio-systems can be profitable (independently from a 
possibility to construct a finer, so to say hidden variables, theory). Yes, I understand that bio-randomness is nonclassical and it is not covered by CP
and hence classical information theory cannot serve as the base for information biology.   But, why do you sell one concrete nonclassical probability theory, namely, 
QP and the corresponding information theory? May be (as Zeilinger guessed) QP and quantum information serve well for a special class of physical systems 
and their application to biology may meet hidden pitfalls?'' Yes, we agree that there is a logical gap: nonclassical does not imply quantum. Negation of 
CP is not QP.

Nevertheless, why exploring QP and quantum information are so attractive?  First of all, because these are the most elaborated nonclassical 
theories.  It is attractive to use their advanced methods. Then one may expect that some basic nonclassicalities of random responses of bio-systems
can be described by the standard quantum formalism. This strategy can be fruitful, especially at the initial stage of development, as it was done 
for quantum cognition \cite{UB}, \cite{11}, \cite{30a}. 

Now the crucial question arises:

\medskip

{\it Can the standard quantum formalism cover completely information processing by bio-systems?}

\medskip

Here ``standard'' refers to Schr\"odinger's equation for the state evolution, the representation of observables by Hermitian operators and 
 the von Neumann-L\"uders projection postulate for quantum measurement.

Our answer is ``no''. Already in quantum information one uses the open system dynamics and generalized observables given by {\it positive operator 
valued measures} (POVMs). It is natural to expect that they can also arise in bio-modeling. Another argument to departure 
from the standard quantum formalism is that in the {\it constructive wave function approach}, i.e., construction of complex probability amplitudes 
from data of any origin violating LTP \cite{UB}, in general observables
are represented not by Hermitian operators, but by  POVMs and even by generalized POVMs which do not sum up to the unit operator \cite{UB}.

A strong argument against the use of solely Hermitian observables in cognitive science was presented in \cite{PLOS}: it seems that the Hermitian description of the 
order effect for a pair of observables $A,B,$ i.e., disagreement between $A-B$ and $B-A$ probabilities,  is incompatible with  $A-B-A$ 
respectability: first $A$-measurement, then $B$-measurement, then again $A$-measurement and the first and the last values of $A$ should 
coincide with probability 1. At the same time the standard opinion polls demonstrating the order effect have the property of  
$A-B-A$ respectability. However, it seems that this problem is even more complicated: even the use of POVMs cannot help so much \cite{PLOS}. 
It seems that we have to go really beyond the quantum measurement formalism, beyond theory of quantum instruments. 
One of such novel generalizations was proposed by the authors \cite{FOOP}, \cite{BOOK} - quantum adaptive dynamics.  
 
\section{Open quantum systems, adaptive dynamics}

The dynamics of an isolated quantum system is described by the Schr\"odinger equation. In the standard quantum framework measurements are mathematically
represented by orthogonal projectors $(P_i)$ onto eigen-subspaces corresponding to the observed values $(a_i)$. A quantum observable can be formally 
represented as a Hermitian operator $A=\sum_i a_i P_i.$ 
The probability to obtain the fixed value $a_i$ as the result of measurement is given by {\it Born's rule}. Let a system have the physical state given by 
a normalized vector $\psi$ of complex Hilbert space $H.$ Then
\begin{equation}
\label{LLBR}
p(a_i)= \langle P_i \psi\vert \psi\rangle= \Vert P_i \psi \Vert^2
\end{equation}
and the post-measurement state is given by 
\begin{equation}
\label{LLBR1}
\psi_{a_i} = P_i \psi/\Vert P_i \psi \Vert. 
\end{equation}

However, the situation that an isolated quantum system propagates in space-time and then 
suddenly meets a measurement device is too ideal. In the real situation a quantum system interacts with other systems, the presence of an environment 
cannot be ignored. In particular, measurement devices can also be treated as special environments.  
The corresponding part of quantum theory is known as theory of {\it open quantum systems.} Here dynamics of a state is described by quantum master equation,
its Markovian approximation is known as  equation  {\it Gorini-Kossakowski-Sudarshan-Lindblad} (GKSL) equation \cite{OV}.\footnote{We remark that quantum master equation (as well as the Schr\"odinger equation) is a linear first order 
(with respect to time) differential equation. Linearity is one of the fundamental features of quantum theory. It is very attractive even from the purely operational
viewpoint, since it simplifies essentially calculations. In fact, mathematically we proceed with matrix-calculus. The question whether QM can be treated as linearization of
more complex nonlinear theory was actively discussed in quantum foundations. For a moment, it is commonly accepted that QM is fundamentally linear, although there were 
presented strong reasons in favor of the linearization hypothesis. This problem is very important for cognition as well. Opposite to physics, even preliminary 
analysis of this problem has not yet been performed.} This equation has numerous applications in quantum 
physics, especially quantum optics. It was applied to model decision making (in games of the Prisoners' Dilemma type) in a series of papers \cite{a1}--\cite{aa2}, 
see also \cite{polina} for applications for decision making in political science. However, the GKSL-equation is an approximation and its derivation is based on a number of assumptions constraining 
essentially the domain of its applications.
One of the main assumptions is that a environment is huge comparing with a system under study. We cannot apply the GKSL-equation to model the state dynamics of 
an electron interacting with another electron or a few electrons considered as an environment.

In decision making typically there are two brain's functions, as, e.g., sensation and perception, which interact and produce the output of one of them, e.g., the output-perception.
In such a situation the GKSL-equation is not applicable. In our work \cite{FOOP} we developed general theory of {\it quantum adaptive dynamics} which, in particular, 
contains the standard 
theory of open quantum systems as describing a special sort of adaptivity, to a bath, see \cite{BOOK}.
In our novel formalism the state dynamics is described by more general class of state transformations than in the standard theory. 
In particular, in theory of open quantum systems
all state transformations are {\it completely positive maps.} An adaptive dynamical map need not be completely positive nor even simply positive. The 
tricky point is that in quantum physics,  for a given state, one can in principle measure any observable. It seems that in  problems of decision making 
this assumption can be relaxed. We consider ``generalized states'' which permit measurements of only special (state-dependent) classes of observables; often  just 
two observables,e.g., sensation and perception. The class of adaptive dynamical maps is essentially larger than the class of (completely) positive maps. This simplify essentially 
modeling of concrete phenomena, for example, recognition of ambiguous figures \cite{ASS}.       

\section*{Appendix}

\subsection*{Quantum nonlocality in physics and biology}  
  
Nonlocal interpretation of violation of Bell's  inequality led to revolutionary rethinking of foundations of QM.  The impossibility to create theories with 
local hidden variables makes quantum randomness even more mystical than it was seen by fathers of QM, for example, by von Neumann \cite{VN}.
In biology we have not yet seen violations of  Bell's  inequality in so to say nonlocal setting. Nevertheless, it  may be useful to comment possible consequences of such possible violations. 
Consider, for example, quantum modeling of  cognition and the issue of mental nonlocality. The quantum-mental analogy has to be used with some reservations. 
The brain is a small physical system (comparing with distances covered by propagating light). Therefore  ``mental-nonlocality'' (restricted to 
information states produced by a single brain) is not as mystical as physical nonlocality of QM. One may expect that in future cognition can be ``explained'', e.g., 
in terms of ``nonlocal'' hidden variables. This position was presented by J. Acacio de Barros   in \cite{Acacio} and it is reasonable. 

``Quantum nonlocality'' is still the subject of hot debates about interpretational and experimental issues. From the interpretation side, the main counter-argument 
against the common nonlocal interpretation is that, in fact,  the main issue is not nonlocality, but contextuality of Bell's experiments \cite{KB1}: s
tatistical data collected in a few (typically four) experimental contexts $C_i$ are embedded in one inequality. However, from the viewpoint of CP 
each context $C_i$ has to be represented by its own probability space. It is not surprising or mystical that such contextual data can violate the 
inequality which was derived for a single probability space \cite{INT}, \cite{UB}. This viewpoint is confirmed by experiments in neutron interferometry \cite{Rauch}
showing a violation of Bell's inequality for a single neutron source, but in multi-context setting. In cognitive psychology a similar experiment, on recognition 
of ambiguous figures, was performed by Conte et al. \cite{22}, see \cite{34} for the theoretical basis of this experiment. 

The experimental situation  in quantum physics is not so simple as it is presented by the majority of writers about violations of Bell's inequality.  There 
are known various loopholes in Bell's experimental setups. A loophole appears in the process of physical realization of the ideal theoretically described 
experimental setup. One suddenly finds that what is possible to do in reality differs essentially from the textbook description. To close each loophole 
needs tremendous efforts and it is  costly. The main problem is to close a few (in future all possible) loopholes in one experiment. Unfortunately, this ``big problem''
of combination of loophole closing is practically ignored. The quantum physical community is factually fine with the situation that different loopholes are closed
in different types of experiments. Of course, for any logically thinking person this situation is totally  unacceptable. The two main loopholes which disturb the 
project of the experimental justification of violation of Bell's inequality are the detection efficiency loophole also known as fair sampling loophole and the 
locality loophole. 

\medskip

{\bf Locality loophole:}
It is very difficult, if possible at all, to remove safely two massive particles (i.e., by escaping decoherence of entanglement) to a distance 
which is sufficiently large to exclude exchange by signals having the velocity of light and modifying the initially prepared correlations. Therefore the locality 
loophole was closed by Weihs et al. \cite{WWW} for photons, massless particles.

\medskip

{\bf Detection efficiency/fair sampling loophole:} Photo-detectors (opposite to detectors for massive particles) have low efficiency, an essential part 
of the population produced by a source of entangled photons is undetected. Thus a kind of post-selection is in charge. In terms of probability spaces, we can say 
that each pair of detectors (by cutting a part of population) produces random output described by its own probability space. Thus we again fall to the multi-contextual
situation.\footnote{As was pointed in the first paragraph of this section, contextuality is present even in the ideal formulation, i.e., 
for detectors having 100\% efficiency. Thus the low efficiency of detection just makes the contextual structure of Bell's experiment more visible.} 
Physicists tried to ``solve'' this problem by proceeding under the assumption of fair sampling, i.e., that the detection selection does not modify 
the statistical features of the initial population. Recently novel photo-detectors of high efficiency (around 98\%) started to be used in Bell's tests.
In 2013 two leading experimental groups closed the detection loophole with the aid of such detectors, in Vienna \cite{V13} (Zeilinger's group) and in Urbana-Champaign \cite{U13}
(Kwiat's group).  

\medskip

Experiments in 2013 led to increase of expectations that finally both locality and detection loopholes would be soon closed in a single experiment.  
However, it seems that the appearance of super-sensitive detectors did not solve the problem completely. Photons disappear not only in the process 
of detection; they disappear in other parts of the experimental scheme: larger distance - more photons disappear. It seems that the real experimental 
situation matches with the prediction of A. Khrennikov and I. Volovich that the locality and detection loopholes  would be never 
closed in a single experiment \cite{KH6} (where by detection efficiency we understand the efficiency of the complete experimental setup):
an analog of Heisenberg's uncertainty relation for these loopholes.

Another problem of the experimental verification of violation of Bell's inequality is that experiments closing one of previously known loopholes 
often suffer of new loopholes which were not present in ``less advanced experiments''.

In general one has not to overestimate the value of Bell's test; other test of nonclassicality, for example, violation of LTP, may be less 
controversial.    

Finally, we make a philosophic remark: Bell's story questions the validity of Popper's principle of falsification of scientific theories. It seems that it is impossible to falsify 
the CP-model of statistical physics in any concrete experiment. It seems that one has to accept that the QP-formalism is useful not because 
it was proven that the CP-formalism is inapplicable, but because the QP-formalism is operationally successful and simple in use. (We remark 
that QP is based solely on linear algebra which much easier mathematically than the measure theory serving as the basis of CP.)

\subsection*{``Exotic quantum-like models: possible usefulness for biological information theory}

Quantum adaptive dynamics, although essentially deviating  from the standard quantum formalism, is still based on
{\it complex Hilbert space.} Coming back to the question posed in section \ref{why?} (``why the quantum formalism?'') we remark that, in principle, there is no 
reason to expect that the biological information theory should be based on the representation of probabilities by complex amplitudes, normalized vectors of 
complex Hilbert space. Therefore one may try to explore models which are quite exotic, even comparing with quantum adaptive dynamics.     

Some of these models are exotic only from the viewpoint of using of rather special mathematical tools; otherwise they arise very naturally from 
the probabilistic viewpoint. For example,  we point to a novel model, so-called {\it hyperbolic quantum mechanics}, which was applied to a series of problems of 
cognition and decision making \cite{UB}. Here, the probability amplitudes are valued not in the field of the complex numbers, but in the algebra of the 
hyperbolic numbers, numbers of the form 
$z=x+jy,$ where $x,y$ are real numbers and $j$ is the generator of the algebra satisfying the equation $j^2=+1.$ Hyperbolic amplitudes describe interference
which is stronger than one given by the complex amplitudes used in QM: the interference term is given by the hyperbolic cosine, opposite to the ordinary trigonometric 
cosine in QM. In \cite{UB} there can be found the classification of various types of (probabilistic) interference exhibited in the form of violation of LTP; here the interference
term  is defined as the magnitude of deviation from LTP. It was found that the mathematical classification leads to only two basic possibilities: either the standard trigonometric
interference or the hyperbolic one. Probabilistic data exhibiting the trigonometric interference can be represented in the complex Hilbert space and the data
with the hyperbolic interference in the hyperbolic Hilbert space.  The third possibility is mixture of the two types of interference terms. This leads to representation of 
probabilities in a kind of Hilbert space over hyper-complex numbers. We remind that in quantum foundations  extended studies were performed to check usefulness of various 
generalizations of the complex Hilbert space formalism;  for example, quaternionic QM or non-Archimedean QM \cite{pi1}. It seems that such models were not so much useful in physics;
in any event foundational studies did not lead to concrete experimental results. However, one might expect that such models, e.g., quanternionic QM, can find applications 
in quantum-like biological information. Possible usefulness of non-Archimedean, especially $p$-adic QM, will be discussed later in relation with unconventional probability 
models.

One of the widely known unconventional probabilistic model is based on relaxation of the assumption that a probability  measure has to value in the segment $[0,1]$  of the real line.
So-called ``negative probabilities'' appear with strange ragularity in a variety of physical problems , see, e.g., \cite{INT}, for very detailed presentation
(we can mention, for examples, the contributions of such leading physicists as P. Dirac, R. Feynman, A. Aspect). Recently negative probabilities started to be used 
in cognitive psychology and decision making \cite{Oas}.

Negative probabilities appears naturally in $p$-adic quantum models as limits of relative frequencies with respect to p-adic topology on the set of rational numbers 
(frequencies are always rational). P-adic statistical stabilization characterizes a new type of randomness which is different from both  classical and quantum 
randomness. P-adic probabilities were used in some biological applications \cite{pia}: cognition, population dynamics, genetics.

\end{document}